\documentclass[conference]{IEEEtran}

\usepackage{cite}
\usepackage{amsmath,amssymb,amsfonts}
\usepackage{algorithmic}
\usepackage{graphicx}
\usepackage{textcomp}
\usepackage{xcolor}
\usepackage{array,multirow}
\usepackage{tablefootnote}
\usepackage{threeparttable}
\usepackage{relsize}
\usepackage{enumitem}
\usepackage{amsmath}
\usepackage{url}
\usepackage{upgreek}
\usepackage[caption=false, font=footnotesize]{subfig}

\def\BibTeX{{\rm B\kern-.05em{\sc i\kern-.025em b}\kern-.08em
    T\kern-.1667em\lower.7ex\hbox{E}\kern-.125emX}}

\makeatletter
\IEEEtriggercmd{\reset@font\normalfont\fontsize{7.9pt}{8.60pt}\selectfont}
\newcommand*\titleheader[1]{\gdef\@titleheader{#1}}
\AtBeginDocument{%
  \let\st@red@title\@title
  \def\@title{%
    \bgroup\normalfont\normalsize\centering\@titleheader\par\egroup
    \vskip1ex\st@red@title}
}
\makeatother
\IEEEtriggeratref{1}

\titleheader{\color{gray}\vspace{-30pt}Presented at the 28th IEEE Int'l. Conference on Electronics, Circuits, and Systems (ICECS), 28/11--01/12 2021, Dubai, UAE.\\This is the authors’ version of the paper. The final published version is posted on IEEE Xplore.\\ DOI: \url{https://doi.org/10.1109/ICECS53924.2021.9665462}}

\title{FPGA \negthinspace \& \negthinspace VPU Co-Processing in Space Applications: Development and Testing with DSP/AI Benchmarks}
    
\newcommand{\ttt}{$\times$}    
    
\makeatletter
\def\ps@IEEEtitlepagestyle{
  \def\@oddfoot{\mycopyrightnotice}
  \def\@evenfoot{}
}
\def\mycopyrightnotice{
  {\footnotesize
  \begin{minipage}{\textwidth}
  \centering\color{gray}%
  ~\copyright~2021 IEEE.  Personal use of this material is permitted.  Permission from IEEE must be obtained for all other uses, in any current or future media, including reprinting/republishing this material for advertising or promotional purposes, creating new collective works, for resale or redistribution\\to servers or lists, or reuse of any copyrighted component of this work in other works.
  \end{minipage}
  }
}
    
\begin{document}

\DeclareRobustCommand{\IEEEauthorrefmark}[1]{\smash{\textsuperscript{\footnotesize #1}}}

\author{\IEEEauthorblockN{Vasileios Leon\IEEEauthorrefmark{1}, 
    Charalampos Bezaitis\IEEEauthorrefmark{2}, 
    George Lentaris\IEEEauthorrefmark{1}, 
    Dimitrios Soudris\IEEEauthorrefmark{1},
    Dionysios Reisis\IEEEauthorrefmark{2},\\
    Elissaios-Alexios Papatheofanous\IEEEauthorrefmark{2},
    Angelos Kyriakos\IEEEauthorrefmark{2},
    Aubrey Dunne\IEEEauthorrefmark{3},
    Arne Samuelsson\IEEEauthorrefmark{4},
    David Steenari\IEEEauthorrefmark{5}}\\[-9pt]
    \IEEEauthorblockA{\IEEEauthorrefmark{1}\emph{National Technical University of Athens, 15780 Athens, Greece}}
    \IEEEauthorblockA{\IEEEauthorrefmark{2}\emph{National and Kapodistrian University of Athens, 15772 Athens, Greece}}
    \IEEEauthorblockA{\IEEEauthorrefmark{3}\emph{Ubotica Technologies Limited, D11KXN4 Dublin, Ireland}}
    \IEEEauthorblockA{\IEEEauthorrefmark{4}\emph{Cobham Gaisler AB, 41119 Gothenburg, Sweden}}
    \IEEEauthorblockA{\IEEEauthorrefmark{5}\emph{European Space Agency, Keplerlaan 1, 2201 AZ Noordwijk, Netherlands}}
}

\maketitle

\begin{abstract}
The advent of computationally demanding algorithms and high data rate instruments in new space applications pushes the space industry to explore disruptive solutions for on-board data processing. We examine heterogeneous computing architectures involving high-performance and low-power commercial SoCs. The current paper implements an FPGA with VPU co-processing architecture utilizing the CIF \& LCD interfaces for I/O data transfers. A Kintex FPGA serves as our framing processor and heritage accelerator, while we offload novel DSP/AI functions to a Myriad2 VPU. We prototype our architecture in the lab to evaluate the interfaces, the FPGA resource utilization, the VPU computational throughput, as well as the entire data handling system's performance, via custom benchmarking.
\end{abstract}

\section{Introduction}

NewSpace trends and the constantly increasing computational demands   
in space applications
mark a new era for on-board data processing.
Existing general-purpose rad-hard CPUs
seem unable to meet future mission requirements
in terms of I/O and computational workloads.
To achieve
high-performance embedded computing 
with enhanced dependability and low power consumption,
the space industry is exploring innovative platforms 
and disruptive avionics architectures.
In this direction,
FPGAs are favored also due to their attractive performance-per-Watt ratio \cite{lentarisTVID}\cite{access}.
However,
recent specialized SoCs, such as the Vision Processing Units (VPUs),
have gained momentum due to their excellence in DSP/AI tasks 
and code development flexibility \cite{leotome}.

The enhanced performance of both aforementioned platforms
can facilitate edge computing in space, 
e.g., for Vision-Based Navigation (VBN) and Earth Observation (EO),
and thus,
avoid the downlink transmission of huge amounts of sensor data to 
the ground stations.
Furthermore,
their I/O flexibility allows servicing multiple instruments/sensors
concurrently,
and complying with multiple protocol specifications.
Towards improved adaptability to various mission scenarios
with minor modifications and in-flight re-programmability,
the space community examines heterogeneous architectures
utilizing distinct types of processors.
When targeting significant improvement 
in performance,
as well as Space, Weight and Power (SWaP) costs,
mixed-criticality architectures are adopted \cite{architectures}\cite{criticality1},
which consist of space-grade and Commercial-Off-The-Self (COTS) components.

The literature includes a number of related works 
\cite{leotome}\cite{gpu_fpga}\cite{gpu_fpga_vpu}.
A co-processing FPGA \& VPU architecture is evaluated in \cite{leotome},
where the VPU is used for accelerating VBN pipelines with limited power consumption.
The hybrid architecture of \cite{gpu_fpga}
bases on the SmartFusion2 SoC FPGA
and the Tegra X2/X2i GPU (main accelerator).
In \cite{gpu_fpga_vpu},
an heterogeneous architecture is proposed,
consisting of 
a SoC FPGA for SpaceWire I/O transcoding,
the AMD SoC (CPU \& GPU) for acceleration,
and optionally a VPU for AI deployment.
When considering CPU \negthinspace \& \negthinspace FPGA co-processing, 
the most prominent devices are the COTS SoC FPGAs,
which integrate both types of processors in a single chip \cite{lentarisTVID} \cite{mpsoc}.

The current paper implements and evaluates a mixed-criticality 
architecture
utilizing an FPGA \negthinspace \& \negthinspace VPU pair in a co-processing configuration. 
The FPGA can be either COTS or space-grade
depending on mission requirements,
and plays the role of the framing processor interfacing with multiple instruments.
Additionally, the FPGA can host high-performance heritage functions, e.g., 
data compression.
The COTS VPU is our key DSP/AI accelerator.
Overall,
the FPGA receives data via on-board sensors or instruments,
e.g., via SpaceWire,
it performs transcoding and forwards the data to VPU via its Camera Interface (CIF).
The VPU acceleration cores execute the DSP/AI algorithm
and return considerable amounts of output data to the FPGA 
via its Liquid Crystal Display (LCD) interface.
We prototype the above architecture 
in the lab 
using the Xilinx XCKU060 FPGA and Intel Movidius Myriad2 VPU 
by developing the CIF and LCD controllers at both sides.
For evaluation,
we implement custom SW benchmarks 
representing common DSP and AI workloads,
which however induce  diverse complexity in I/O, computation, and storage.
We assess the functionality and speed of the interfaces,
the FPGA resource utilization,
the VPU processing capabilities,
and the performance of the entire system (I/O \& processing).
The results show that 
our FPGA \negthinspace \& \negthinspace VPU co-processors achieve 
6--20 FPS
for kernels such as binning, convolution, and rendering,
as well as more than 1 FPS for deep AI image classification on 1MPixel   images.
The FPGA resource utilization is limited 
and leaves room for extra HDL components,
such as I/O transcoding
and data compressors,
or accelerators of more demanding algorithms.

The remainder of the paper is organized as follows. 
Section II introduces the proposed architecture.
Section III includes development details on the FPGA and VPU regarding the I/O interfaces and the benchmarks.
Section IV discusses the evaluation results.
Finally, Section V draws the conclusions. 

\section{Proposed Architecture}

\begin{figure}
    \centering
  \hspace{-10pt}  \includegraphics[width=1.02\columnwidth]{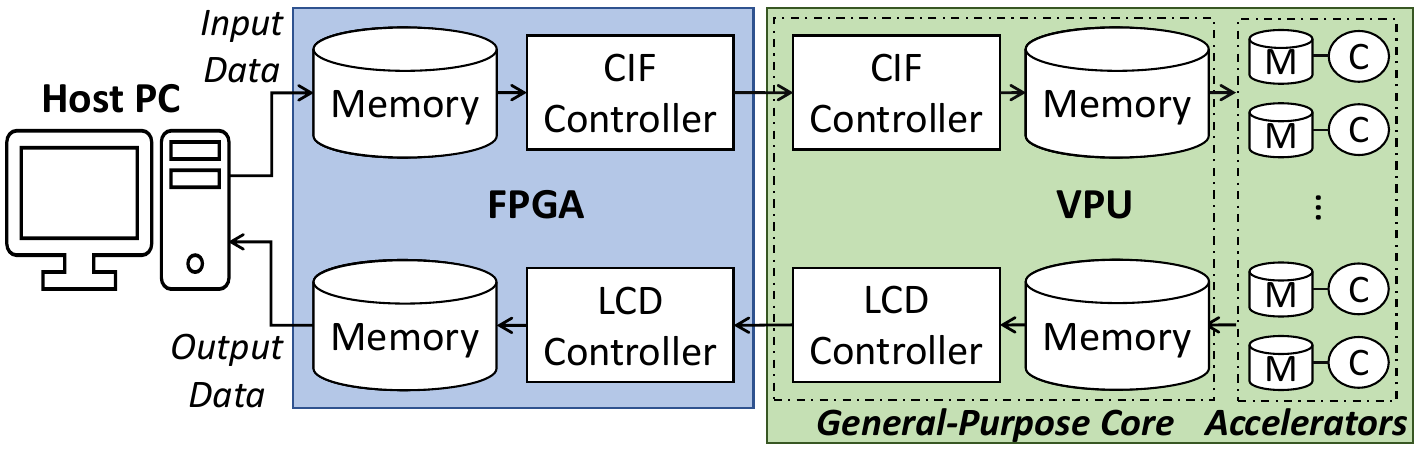}\\[-5pt]
    \caption{Testbed architecture: FPGA \& VPU co-processor with CIF/LCD I/O.}
    \label{archi}
    \vspace{-7pt}
\end{figure}

Our proof-of-concept architecture for FPGA \negthinspace \& \negthinspace VPU co-processing
is illustrated in Fig. \ref{archi}.
At the FPGA side,
we implement in VHDL the controllers for both CIF and LCD, 
as well as buffers for storing the data.
Similarly, 
at the VPU side, 
we call the vendor's routines for handling the CIF/LCD interfaces,
store the data in the DRAM memory,
and parallelize our algorithms on the acceleration cores.
Our Host PC is responsible for
transferring the I/O data to/from the FPGA
and validating the results via comparisons to groundtruth data.

The CIF/LCD interface modules include various signals, e.g., for 
clocks (\emph{pixel\_clock}), 
data (\emph{pixel}),
and synchronization (\emph{hsync}, \emph{vsync}).
Each module integrates several components, 
such as 
control and status registers,
pixel FIFOs, and
encoders/decoders.
Their pixel bit-width (8/16/24) and the clock frequency 
can be configured independently. 
Based on our preliminary tests,
both interfaces will operate at up to 50MHz without errors.
This frequency is expected to provide an I/O throughput of up to 50 pixels per $\upmu$s, 
i.e., for example, transmit a 1024\ttt1024 frame in 20.9ms.

The main accelerator of our architecture is a VPU \cite{myriad, gap8},
which is a prominent SoC for embedded DSP and AI applications,
consuming \raisebox{0.8pt}{$\scriptstyle\sim$}1W.
A single VPU chip integrates numerous heterogeneous processors,
i.e.,
general-purpose cores, vector cores, HW filters, and AI engines,
allowing the efficient deployment of complex workloads.
In addition,
it provides various memories (DRAM, scratchpad, caches)
and HW peripherals (e.g., CIF, LCD, SPI).

Our proof-of-concept implementation is also ported to the High-Performance Compute Board (HPCB) \cite{joaquin},
where Xilinx Kintex XCKU060 \cite{kintex_link} is the FPGA 
(to be replaced with space-grade XQRKU060 for missions),
and Intel Movidius Myriad2 \cite{myriad_link} is the VPU.
The architecture includes a radiation-tolerant microcontroller (Cobham Gaisler GR716 \cite{gr716}),
which is the reliable supervisor of the FPGA \negthinspace \& \negthinspace VPU co-processor.
The main interfaces of HPCB are 
4 SpaceFibre (3.1-6.3 Gbps) and 2 SpaceWire (100 Mbps) links.
Moreover, 
to provide fault-tolerance and/or increased performance,
the platform includes 3 VPUs.
HPCB can be integrated in 
the On-Board Computer (OBC) 
and/or 
the payload data handling unit
of future spacecrafts,
providing 
simultaneous processing of multiple instruments
and high-performance DSP/AI.

The current paper focuses on evaluating 
the potential performance of our FPGA \negthinspace \& \negthinspace VPU co-processing scheme 
by using Virtex-7 XC7VX485T (and XCKU060) as FPGA and Myriad2 as VPU
to assess data transfers and custom acceleration.
Our future work includes product verification on HPCB boards \cite{joaquin}.

\section{Development on FPGA and VPU}

\subsection{I/O Interfaces \& Dataflow of FPGA}

\begin{figure*}
    \centering
  \includegraphics[width=0.55\textwidth]{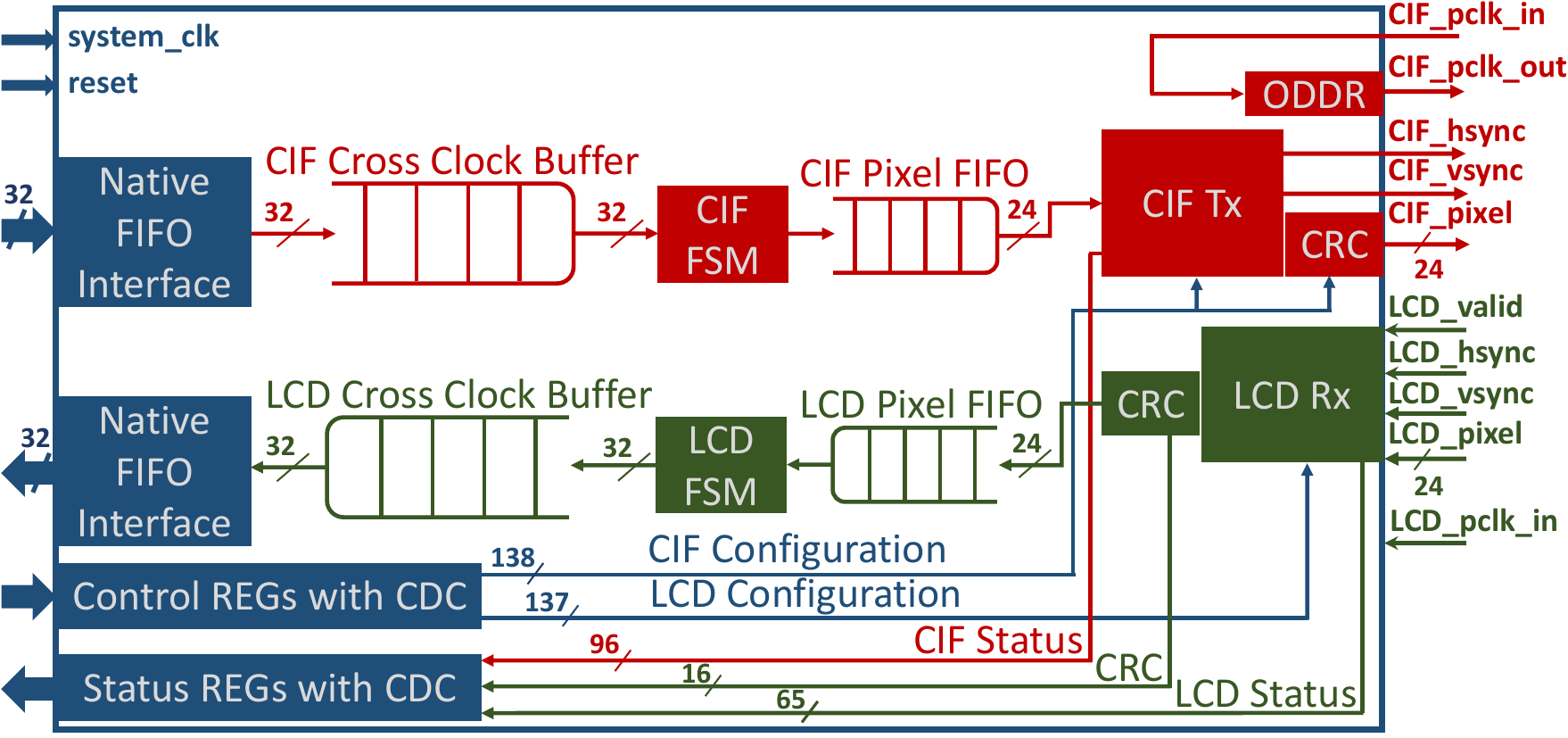}\\[-5pt]
    \caption{Block diagram of CIF/LCD I/O interface in FPGA.}
    \label{cif_lcd}
    \vspace{-7pt}
\end{figure*}

The CIF module of the FPGA,
which injects data into the VPU,
is depicted in Fig. \ref{cif_lcd}.
The CIF image buffer is implemented with a native FIFO interface 
connecting CIF with the FPGA internal bus.
CIF waits for data bursts
to be stored in the image buffer,
and then forwards them in the FSM stage.
The CIF FSM converts the 32-bit input data 
to the CIF pixel bit-width (8/16/24),
and writes them to the pixel FIFO.
The CIF Tx component,
which implements the CIF protocol
and handles the \emph{hsync} and \emph{vsync} synchronization signals,
reads the pixel FIFO and controls the data transmission to the VPU.
In the end of the CIF dataflow, 
a Cyclic Redundancy Check (CRC) component appends the calculated CRC-16/XMODEM 
to the last line of the frame to be transmitted.
 
Fig. \ref{cif_lcd} also presents the 
LCD module of the FPGA,
which handles the data received from the VPU.
LCD Rx is the first component in the image reception dataflow.
It writes one pixel per clock cycle in the LCD pixel FIFO based on the synchronization signals (\emph{hsync}, \emph{vsync}),
which are
generated by the VPU.
Similarly to CIF,
the LCD FSM converts the LCD pixel formats (8/16/24) to 32-bit words
and writes them in the LCD image buffer to be forwarded to the FPGA bus.

Our design also includes control registers for both CIF and LCD,
which are written at runtime to configure the frame dimensions and the pixel bit-width.
Moreover,
status registers are updated at runtime
when an input/output frame is transmitted/received.
These registers 
store information,
such as the CRC results of both directions
and the total number of frames transmitted/received,
and report it
to the system's control.
It is worth mentioning  
that our FPGA design uses
FIFOs capable of clock domain crossing,
allowing different clocks to be employed
for the CIF and LCD modules.

\subsection{I/O Interfaces \& Dataflow of VPU}

\begin{figure}
    \centering
    \hspace*{-14pt} \includegraphics[width=0.55\textwidth]{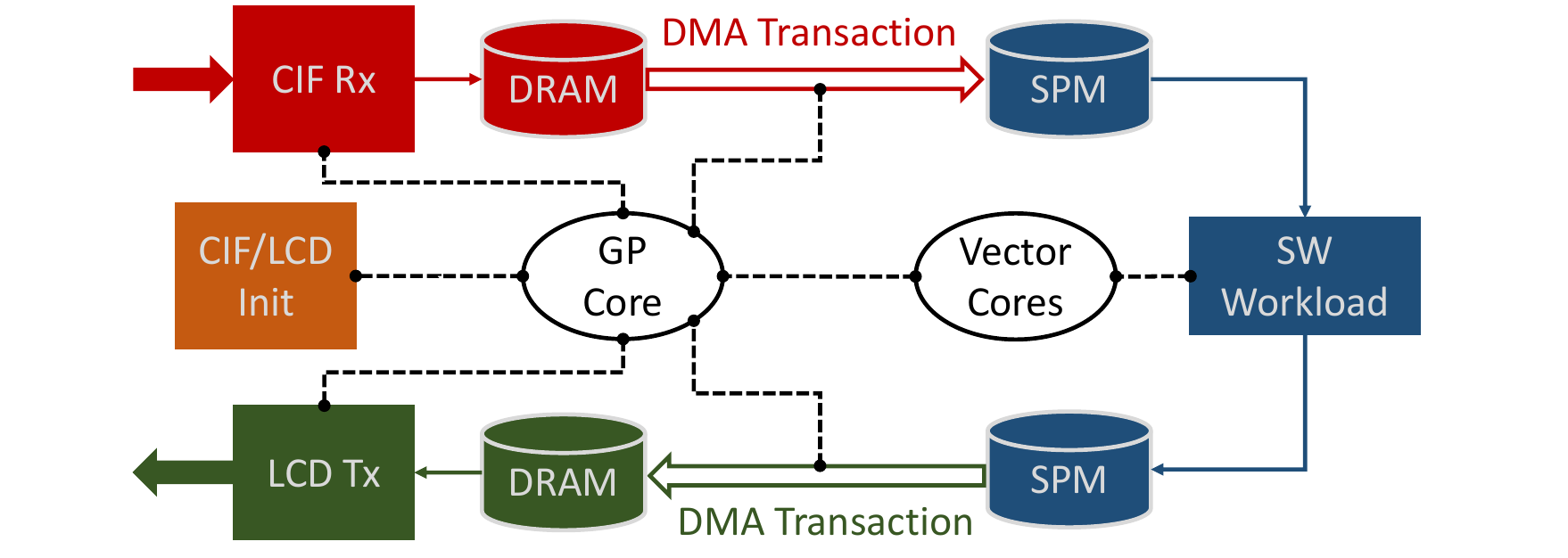}\\[-5pt]
    \caption{Dataflow in VPU: CIF/LCD I/O, memory transactions and processing.}
    \label{myriad}
    \vspace*{-7pt}
\end{figure}

The SW environment of the Myriad2 VPU 
includes the \emph{CamGeneric} component for handling CIF,
which abstracts multiple driver layers.
This component includes all the SW routines
for operating the HW CIF engine.
We target the parallel bus of CIF, 
and call routines such as \emph{CamInit()}, \emph{CamStart()} and \emph{CamStop()}.
Similarly for LCD,
the SW environment provides 
a library of routines, such as \emph{LCDInit()}, \emph{LCDQueueFrame()}, \emph{LCDStartOneShot()} and \emph{LCDStop()}.

The dataflow in the Myriad2 VPU is presented in Fig. \ref{myriad}.
To establish the CIF/LCD communication,
the General-Purpose (GP) LEON processor configures the corresponding GPIOs,
prepares the driver's settings (e.g., frame size, camera buffers),
and
initializes the HW engines via the \emph{CIFInit()} and \emph{LCDInit()} routines.
The CIF Rx function
receives the frame 
by calling the necessary routines,
and stores it in DRAM.
To start processing,
LEON initializes
the main acceleration processors, 
i.e., the 12 SHAVE cores (VLIW \negthinspace \& \negthinspace SIMD, 600MHz, bare-metal),
and transmits the input frame to their Scratchpad Memory (SPM)
via the DMA engine.
The output data of the processing
are transmitted from SPM to DRAM,
again via the DMA engine,
and then,
our LCD Tx function calls the corresponding LCD routines 
to transmit them to FPGA.

\subsection{Custom SW Benchmarks}

The \emph{Averaging Binning} benchmark works on 2\ttt2 image regions with stride 2 to calculate their mean value.
We divide the 2048\ttt2048 8-bit input image into 36 bands,
and each SHAVE is assigned 3 bands to process. 
The processing is performed in-place on the input buffer.
Moreover, we have enabled the cache of SHAVEs.
A similar design approach has been adopted for our second benchmark, 
i.e., \emph{Floating-Point (FP) Convolution},
which processes 1024\ttt1024 8-bit input images
with different kernel sizes (3\ttt3--13\ttt13).  

\emph{Depth Rendering}
uses a triangle mesh model and a 6D vector
to generate a 1024\ttt1024 16-bit image
with pixels encoding the distance between camera
and the nearest point on the model's surface.
The processing is based on rasterization,
and involves tasks such as 
triangle projection, 
bounding box traversal,
and
distance calculation.
To decrease the idle time, 
each SHAVE is dynamically assigned a new band to render, 
upon finishing its previous one.
To provide extra acceleration,
SHAVEs use SIMD routines for the main computations.
Regarding memory,
we use only one working buffer (Z-buffer) in CMX,
and also, 
store the static model in DRAM
to give SHAVEs access to the entire model.

Our last benchmark is \emph{CNN Ship Detection},
which implements a 6-layer network (132K parameters)
to detect ships on 1024\ttt1024 16-bit RGB satellite images.
The training is performed with TensorFlow on 128\ttt128 RGB images 
and the classification accuracy is 96.8\% \cite{kaggle}.
For our implementation, 
the 32-bit FP weights and input images are converted to 16-bit FP using the available Myriad2 functions.
The inference engine is implemented on SHAVEs for 128\ttt128 images.
Therefore,
a function running on LEON divides the 1024\ttt1024 input image into 64 patches, 
stores each patch in the engine's input buffer (placed in DRAM), 
and then orders the SHAVEs to start the patch processing.

\section{Evaluation}

\begin{figure}
    \centering
  \subfloat[GR-VPX-XCKU060]{%
    \includegraphics[width=0.70\columnwidth]{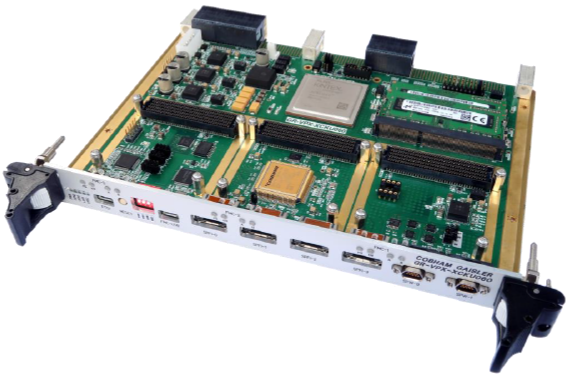}}\\
  \subfloat[GR-HPCB-FMC-M2]{%
    \includegraphics[width=0.43\columnwidth]{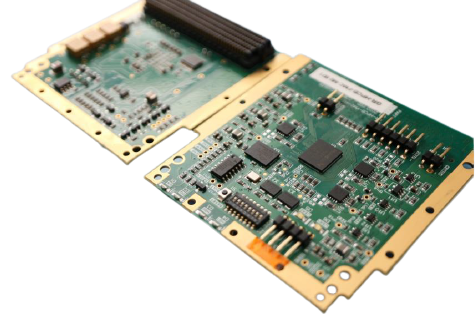}}
  \caption{HPCB platform: (a) FPGA motherboard and (b) VPU mezzanine \cite{joaquin}.}
  \label{boards} 
\end{figure}

Initially,
we examined the correct functionality of our CIF/LCD interface,
using a loopback scenario 
and testing different CIF/LCD frequencies
on our XC7VX485T--Myriad2 setup.
The tests were also ported to the HPCB platform (depicted in Fig. \ref{boards}),
which consists of a motherboard hosting the XCKU060 FPGA 
and an FMC mezzanine card hosting the Myriad2 chip.
We successfully performed data transmission without errors
for 8-bit 2048\ttt2048 frames
at 50MHz.
Due to the FPGA memory resources,
we transmitted without errors 16-bit frames 
with up to 1024\ttt1024 size.
To achieve higher frequencies for CIF and LCD, e.g., 100MHz,
we had to decrease the FPGA buffers.
Thus,
we successfully trasmitted 
up to 16-bit 64\ttt64 frames
with
CIF operating at 100MHz 
and LCD operating at 90MHz.

Table \ref{fpga} reports the resource utilization 
of the CIF/LCD interface implementation on XCKU060.
Our design utilizes $<$1\% of the total chip's resources for every FPGA primitive
(3.5K LUTs, 1.6K DFFs, 7 DSPs, 6 RAMBs),
to support the transmission of up to 4MPixel 24-bit frames.
In our co-processing architecture,
the FPGA can also
execute/accelerate other tasks,
such as 
I/O instrument transcoding, 
data compression, 
and
heritage DSP functions.
In this direction,
we report implementation results 
for 
\emph{CCSDS-123.0-B-1} \cite{paschalis} (hyperspectral image compression),
\emph{FIR Filter} (signal processing)
and
\emph{Harris Corner Detector} (image feature detection). 
The hyperspectral compression (AVIRIS sensor, parallelization=1) 
consumes 11\% LUTs and 6\% DFFs,
while the corresponding utilization for our DSP functions is below 2\%.
In terms of memory,
the compression (680\ttt512\ttt224 image) 
and corner detection (1024\ttt32 image band) 
kernels
consume \raisebox{0.8pt}{$\scriptstyle\sim$}6\% RAMBs.
Hence,
the FPGA leaves room for integrating additional HDL components
of algorithmic pipelines.

\begin{table}[!t]
\renewcommand{\arraystretch}{1.21}
\setlength{\tabcolsep}{2.7pt}
\caption{Resources of FPGA$^1$ as Framing Processor \& Accelerator}
\vspace{-4pt}
\label{fpga}
\centering
\begin{threeparttable}
\centering
\begin{tabular}{l|c|cccc}
\hline
\multicolumn{1}{c|}{\bf{Design}} & \bf{Parameters} & \textbf{LUT} & \textbf{DFF} & \textbf{DSP} & \textbf{RAMB} \\
\hline
\hline
CIF/LCD Interface        & 4MP, 24bpp   & 1\% & 0.3\% & 0.3\% & 0.6\%  \\
CCSDS-123 \cite{paschalis}$^2$  & 680\ttt512\ttt224, 16bpp   & 11\% & 6\% & 0.2\% & 6\% \\
FIR Filter    & 64-tap, 16bpp  & 0.5\% & 0.5\% & 2\% & 0\% \\
Harris Corner Detect.      & 1024\ttt32, 8/32bpp  & 2\% & 2\% & 2\% & 6\% \\
\hline
\end{tabular}
\begin{tablenotes}
   \item[1] {\scriptsize  Kintex UltraScale XCKU060 (331K LUTs, 663K DFFs, 2.7K DSPs, 1K RAMBs).} 
   \item[2] {\scriptsize resource utilization extrapolated from the Kintex-7 XC7Z045 implementation of \cite{paschalis}. }
\end{tablenotes}
\end{threeparttable}
\vspace{-5pt}
\end{table}

Next,
we assess the acceleration of our benchmarks using the general-purpose LEON processor as baseline.
For \emph{Averaging Binning}, 
we achieve 14$\times$ speedup,
which mainly comes from the parallelization to 12 cores
(LEON has to scan the entire 4MP image, adding significant timing overhead).
Depending on the kernel size,
we achieve up to 75$\times$ speedup in our \emph{FP Convolution} benchmark. 
Compared to \emph{Averaging Binning},
the speedup is larger due to increased computational complexity 
(according to our measurements,
the processing power of LEON is almost equal to that of 2 SHAVEs).
Regarding \emph{Depth Rendering},
the speedup is 10--16$\times$ depending on the input image,
as the performance of this workload is highly associated with the content of the image.
Finally,
\emph{CNN Ship Detection} was implemented only on SHAVEs,
however,
considering the performance of convolutions,
and given that LEON does not support 16-bit FP
(thus it will execute the 32-bit FP model),
the speedup is expected to be more than 2 orders of magnitude.

\begin{table*}[!t]
\renewcommand{\arraystretch}{1.21}
\setlength{\tabcolsep}{3.8pt}
\caption{Evaluation Results of the Proposed FPGA \& VPU Co-Processing Architecture with CIF/LCD@50MHz I/O Interfaces}
\vspace{-4pt}
\centering
\resizebox{\textwidth}{!}{  
\begin{threeparttable}
\centering
\begin{tabular}{l|c|ccc|cc|cc}
\hline
\multicolumn{1}{c|}{\multirow{2}{*}{\bf{Benchmark}}} & \multirow{2}{*}{\bf{I/O Data}} & \bf{CIF Input} & \bf{VPU Processing} & \bf{LCD Output}  & \multicolumn{2}{c|}{\bf{Unmasked I/O$^1$}}   &  \multicolumn{2}{c}{\bf{Masked I/O$^2$}}     \\[-2pt]
 & & \bf{Time} & \bf{Time}  & \bf{Time}  &  \bf{Latency} & \bf{Throughput} & \bf{Latency} & \bf{Throughput} \\
\hline
\hline 
Averaging Binning & 4MP/1MP, 8bpp& 85ms	& 3ms	& 21ms & 109ms & 9.1 FPS & 906ms & 3.2 FPS  \\
3\ttt3 FP Convolution&  1MP/1MP, 8bpp & 21ms & 8ms & 21ms & 50ms & 20 FPS & 336ms & 8 FPS   \\
7\ttt7 FP Convolution & 1MP/1MP, 8bpp & 21ms & 29ms & 21ms & 71ms & 14.1 FPS & 336ms & 8 FPS   \\
13\ttt13 FP Convolution & 1MP/1MP, 8bpp & 21ms & 114ms & 21ms & 156ms & 6.4 FPS & 336ms & 8 FPS  \\
Depth Rendering & 6\ttt1/1MP, 16bpp & $<$1$\upmu$s     & 164ms & 21ms & 185ms & 5.4 FPS & 391ms & 6.1 FPS  \\
CNN Ship Detection & 1MP RGB/64\ttt1, 16bpp & 63ms & 658ms & $<$1$\upmu$s & 721ms & 1.4 FPS & 1505ms & 1.5 FPS  \\
\hline
\end{tabular}
\begin{tablenotes}
    \item[*]  {\scriptsize  \textbf{Unmasked I/O} mode assumes serial I/O--processing and  \textbf{Masked I/O} mode assumes pipelined I/O--processing (for streaming processing)}
   \item[1] {\scriptsize  \textbf{Latency} = CIF Time + VPU Time + LCD Time, 
                           \textbf{Throughput} = 1/(CIF Time + VPU Time + LCD Time)}
   \item[2] {\scriptsize  \textbf{Latency} =   $\max\{$VPU Time - LCD Buff. Time, CIF Time + CIF Buff. Time + LCD Time$\}$ + $\max\{$VPU Time, LCD Buff. Time + CIF Time + CIF Buff. Time + LCD Time$\}$ +  LCD Buff. Time + CIF Time + CIF Buff. Time + LCD Time,  \textbf{Throughput} =     1/($\max\{$VPU Time, LCD Buff. Time + CIF Time + CIF Buff. Time + LCD Time$\}$)}
\end{tablenotes}
\end{threeparttable}
\label{results}
}
 \vspace{-6pt}
\end{table*}

The performance of the entire architecture,
involving both I/O and processing, is evaluated using two distinct scenarios: 
\begin{enumerate}[ wide = 1pt, leftmargin = *]
    \item[1)] \underline{Unmasked I/O}: 
    assuming serial I/O--processing,
    the VPU receives the input frame from the FPGA, performs the processing, and transmits the output data to the FPGA.
    \item[2)] \underline{Masked I/O}: 
    assuming pipelined I/O--processing and streaming input, 
    the VPU performs in parallel 2 processes: 
    i) buffering of output frame \emph{n}-1, CIF reception and buffering of input frame \emph{n}+1, LCD transmission of output frame \emph{n}-1, 
    and  
    ii) processing of frame \emph{n}.
    In that case, the one LEON processor of the VPU handles the I/O (process i), and the other manages the processing performed by the SHAVEs (process ii).
 \end{enumerate}
 
The performance results are presented in Table \ref{results}.
Both I/O interfaces are operating at 50MHz and, as expected, transmit an 1MPixel image in \raisebox{0.8pt}{$\scriptstyle\sim$}21ms.
In the Unmasked I/O mode, 
the total throughput ranges between 9--20 FPS for benchmarks with small processing time. 
To implement the Masked I/O mode, 
the input/output data are buffered to an allocated DRAM space for data integrity reasons 
(copying an 1MPixel   frame requires \raisebox{0.8pt}{$\scriptstyle\sim$}42ms).
As a result, 
the latency of a single frame increases considerably.
Even so, those benchmarks featuring excessive processing time can benefit
from this masking technique 
and improve their throughput by 1.1--1.3$\times$.
Table \ref{results} shows this effect on
13\ttt13 \emph{FP Convolution}, \emph{Depth Rendering} and \emph{CNN Ship Detection}. 
In contrast, 
benchmarks with small processing time suffer 
a throughput decrease when applying masking
and the developer must be cautious with respect to the selected mode of operation
(in Table \ref{results}, 
\emph{Averaging Binning} has only 3ms computation
and the buffering of its 4MP input frame adds considerable overhead).

\begin{figure}
    \centering
 \hspace{-9pt}   \includegraphics[width=\columnwidth]{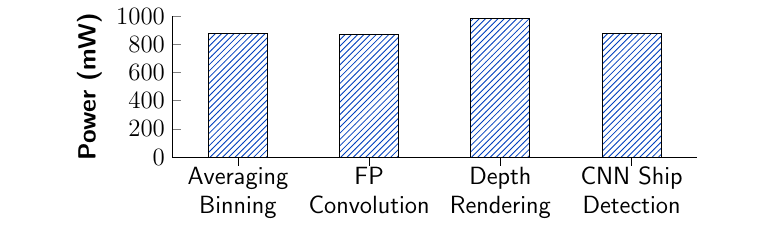}\\[-7pt]
    \caption{Power consumption of the VPU per benchmark execution.}
    \label{power}
    \vspace{-4pt}
\end{figure}

Regarding power consumption, 
the VPU consumes 0.8--1W for all the benchmarks when utilizing SHAVEs,
as shown in Fig. \ref{power}.
The corresponding values for the benchmark implementations on the GP LEON processor lie between 0.6W and 0.7W, 
however,
in terms of FPS/W,
the SHAVE implementations outperform LEON,
e.g., by 11$\times$ for \emph{Averaging Binning} and up to 58$\times$ for \emph{FP Convolution}. 
Moreover, compared to the Zynq FPGA \cite{lentarisTVID}\cite{chatzi},
our implementations on the VPU's SHAVEs deliver 3--4$\times$ better consumption.

When compared to other competitive embedded devices, 
our VPU accelerator
provides \raisebox{0.8pt}{$\scriptstyle\sim$}2.5$\times$ less FPS/W 
vs. the Zynq-7020 FPGA for \emph{CNN Ship Detection} with the same model \cite{chatzi},
however, 
the latter consumes almost all the chip resources.
Therefore,
it would require dynamic reconfiguration to deploy other algorithms,
e.g., 
in case of space applications involving deep DSP/AI pipelines,
adding significant timing penalties.
In contrast,
the VPU SoC can store multiple programs in DRAM and seamlessly execute them at runtime.
Compared to the Jetson Nano GPU \cite{chatzi},
the CNN implementation in VPU delivers \raisebox{0.8pt}{$\scriptstyle\sim$}4$\times$ better FPS/W.
For \emph{Averaging Binning},
our VPU implementation provides \raisebox{0.8pt}{$\scriptstyle\sim$}3$\times$ better throughput than 
a typical Zynq FPGA implementation with 1 binning pipeline on programmable logic (1 input pixel per cycle),
also due to the slower DMA engines of the Zynq SoC.

\section{Conclusion and Future Work}

We proposed and developed an FPGA \& VPU co-processing architecture
to accelerate demanding DPS/AI workloads for space applications.
The results showed that our CIF/LCD interconnection scheme
achieves sufficient I/O throughput,
e.g., 48 FPS for 1MPixel image transfers between FPGA and VPU,
with zero errors.
In terms of computational performance,
we accelerated a variety of image processing tasks,
e.g., binning, rendering, and CNN,
by 1--2 orders of magnitude versus the LEON4 CPU.
At system level,
we increased the throughput rates 
to 6--20 FPS
for small kernels commonly used in VBN pipelines.
For more complex AI benchmarks,
such as CNN classification on 1MPixel images,
we achieved over 1 FPS.
Our future work will focus on 
more extensive evaluation and verification of our techniques 
on the HPCB board/product 
including the integration of multiple
SpaceWire \& SpaceFibre links,
high-level frameworks for SW deployment,
as well as multiple VPUs to improve fault-tolerance and/or throughput. 

\section*{Acknowledgment}
This work was partially supported by the European Space Agency
via the ``FPGA Accelerated DSP Payload Data Processor Board'' activity 
under Grant 4000126129/18/NL/AF.

\bibliographystyle{IEEEtran}
\bibliography{REFERENCES.bib}

\end{document}